# Toulouse 2D numerical radiative transfer codes

[ https://idoc.ias.u-psud.fr/MEDOC/Radiative%20transfer%20codes/MALI-GS-2D ]

**Frédéric Paletou, Ludovick Léger, Martine Chane-Yook**
(march 2019)

We propose two numerical codes for 2D **multilevel non-LTE** radiative transfer, assuming **complete** frequency redistribution, for **static**, **homogeneous** and **isolated** structures. They were initially developed for solar prominence radiative modeling of various spectral lines of H or He.

Two versions are proposed, respectively using the following numerical schemes:

- Multilevel Accelerated Lambda Iteration (MALI),
- A Gauss-Seidel iterative scheme in a 2D cartesian grid.

MALI was first proposed in 1991, and used very quickly after for prominences modelling by Heinzel (1D) and Paletou (2D), both in 1995. This is basically an extension for multilevel atoms of the Accelerated Lambda Iteration (ALI) scheme, whose most efficient version using a diagonal operator (i.e., a Jacobi-type iterative scheme) was first exposed in the seminal article of Olson, Auer and Buchler (1986). However Heinzel's and Paletou's 1995 papers respectively describe how to implement ionization equilibrium, and how to add also partial frequency redistribution effets for resonance lines (which is however *not* included in this distribution).

The 2D formal solver, using the so-called short characteristics methods, was upgraded and described in many details in Auer & Paletou (1994).

The Gauss-Seidel (GS) iterative scheme was first proposed in 1995 by Fabiani Bendicho and Trujillo Bueno, for a 1D two-level atom model. Lambert et al. (2016) propose more explanations about these various iterative schemes.

Paletou & Léger (2007) gave further details (and pointed at a small correction to make after Auer & Paletou 1994), about the implementation of such a scheme for multilevel atoms. One may also find in the later article, updated data which may be very useful for tests with simple multilevel atoms. Indeed, simplified model atoms of H, Ca or Na are proposed hereafter with our codes.

The more tedious implementation of both GS and multilevel atoms in 2D cartesian geometry is well described in Léger et al. (2007). It includes also parts about the implementation of multi-grid methods, which are **not** included in this distribution. *However, an enlightning description of these methods was given (in French, though) in the PhD thesis of Ludovick Léger (2008, to be found at* https://tel.archives-ouvertes.fr/tel-00332781/fr/ *on-line).*

Finally, the interested reader and/or user is recommended to go through Léger & Paletou (2009) for elements about the most detailed He modeling so far in multi-dimensional geometry, including the atomic fine structure, as well as first attempts towards multi-D thread models, and direct comparisons with observed data taken in the visible and, simultaneously, in the near infrared using the spectropolarimetric observing mode of the *Thémis* ground-based solar telescope.

# MALI version

All codes provided hereafter were written in **Fortan 90**.

`atomCaI5n` : data for a schematic 5-level Ca I (see also Paletou & Léger 2007)

`atomHI3n` : same for 3-level H I (ibid.)

`atomNaI4n` : same for 4-level Na I (ibid.)

`boltzex` : computes density of populations according to the Boltzmann distribution

`dopwidth` : computes a Doppler width (a so-called microturbulence velocity may be included)

`eincoef` : computes all Einstein coefficients, per transition

`lubksb`, `ludcmp` : standard procedures for the so-called LU-decomposition

`mali2d` : main programme

`malieqstat` : solves for the preconditioned (linear) system of statistical equilibrium equations (hereafter **EES**), according to Rybicki & Hummer (1991)

`planckf` : computes Planck functions

`ratio` : computes the step-size ratio for logarithmic spatial grids (from L.H. Auer, LANL)

`rt2d` : the so-called formal solver (see details in: Auer & Paletou 1994, Léger et al. 2007)

`seta` : defines angular quadrature for the formal solver (Auer & Paletou 1994, see also the older Mihalas et al. 1978 for a more precise reference to 'sets' A and B)

`setgeo` : computes (once) all sets of points and interpolation weights required by the formal solver rt2d; it is important that the 2D grid if further swept according to the order defined during this first pass (ibid. and details in §3. of Léger et al. 2007 - **critical** for the Gauss-Seidel implementation, hereafter); all relevant data, that will be very often used during each run of the codes, is stored in **tab_wts**.

Subroutine **setgeo** is likely to be the **most critical part** of the whole codes. Any serious user for 2- or 3D RT should first understand with maximum care how this works.

`down` (on top *and* sides), `botup` and `sideup` are used to set the **external** radiation hitting the faces of the 2D slab. Be careful with proper/eventual 'dilution factors' (e.g., Paletou 1996).

Given this set of fundamental subroutines, it may be somewhat tedious, but not difficult to implement ionization equilibrium and a self-consistent determination of the electronic density (see Heinzel and/or Paletou 1995).

**GS2D**: the Gauss-Seidel 2D formal solver version

Most of the material is similar to the one already commented for MALI in 2D. New material, specific to GS iterations, appear hereafter:

`gsm2d` : main programme

`gsmeqstat` : solves for the system of EES, in the frame of the Gauss-Seidel iterative scheme

`rt2dgsm` : is the modified formal solver, required for Gauss-Seidel iterations in 2D

`rt2dgsm` is the **very critical part** for understanding of works an efficient Gauss-Seidel formal solver in 2D (and further). **Any serious user/developer should invest in every details of Paletou & Léger 2007, and Léger et al. 2007 articles before using or even modify our codes.**

The tools and (Python) codes given along with Lambert et al. (2016) may also help for the understanding of such a numerical scheme, including Successive Over-Relaxation (SOR); see the webservice and associated ressources http://rttools.irap.omp.eu/

### Input data

See *ascii* file **input** which carries the following content:

- number of (reduced) **frequencies** (for *half-profile*, may be modified)
- number of **directions** (see **seta**)
- **max**. optical depth, **first** optical depth step, **number** of points (**first in Z, then in Y**)
- grid may be read from extra file (see **lecfic ; MALI only**)
- maximum number of iterations (**niter**)
- (uniform) slab temperature (**ta**)
- model-atom: H or Ca or Na (see *extra-files*: **atomH**, Ca, Na…; see relevant data in Table 2 of Paletou & Léger 2007)
- background continuum opacity (**chic**)
- accelerate convergence after **iaccel** iterations (**MALI only**)
- expected relative error **eps_lu**

**Warning**: for the **GS2D** code, the two last lines/parameters are *replaced* by option SOR (**sor_lu**); and see further parameter **sor** computed in **gsm2d.f90**

**External illumination conditions** (specific intensity) are implemented in respectives variables **down** (all surfaces), **botup** and **sideup**. Be careful with eventual *dilution* factors to take into account (see Paletou 1996).

### Default output data

Two files are used, but of course any other relevant data (emergent intensities, scattering integrals, full 2D distribution of populations, source functions, etc.) can be downloaded.

**population2d.res** : density of populations in 2D obtained after convergence.

**populations.res** : density of populations along vertical axis of symmetry, in order to compare with 1D results (see "Test data" description hereafter).

**mali2d.res** : provides elements about the convergence of the process (e.g., relative errors on populations, or source function, or vs. 1D computations, if necessary/relevant - see hereafter)

The **Gauss-Seidel 2D** code provides only one output file, called either **gsm2d.res** or **sor2d.res** (depending on the selected option in **input**), which contains: the density of populations, then the maximum relative error (from an iteration to another) on the source function, and on populations.

## Test data

For the respective model-atoms of H, Ca and Na, .res (1D) and .save (2D) data can be re-used, for benchmarks, or stay unused, after minor modifications in the source code (where reads are done). 2D data were computed using a large enough horizontal optical thickness so that data along the vertical axis of symetry can be reasonably compared to 1D results.

## Description of (other) major variables used in the code

**phi** and **wtnu** : **absorption line profile array,** and **frequency weights** for numerical integration vs. (reduced) frequencies**.**

**s** and **sl : total source function,** and **line source function.**

**nprec :** used for storage of **data at step (n-1)**, for relative error computations**.**

**chil21, chil** and **chic : Lyman-alpha line opacity distribution, other line opacities,** and **background (fixed) opacity.**

**jeff** and **lstar : effective scattering integral** (see MALI original article), and **diagonal operator.**

## Miscellaneous comments

No **microturbulent** velocity is considered, although it is easy to add into **dopwidth** allocation, if necessary.

**Ionization** equilibrium could also be implemented further, following Paletou (1995) and Heinzel (1995) numerical schemes.

**Partial** frequency redistribution effects could be implemented, following either Paletou (1995) or Uitenbroek (2001, http://cdsads.u-strasbg.fr/abs/2001ApJ...557..389U).

**Moving** structures could also be considered and modeled, after modifications on (i) the frequency quadrature (then over a *full* profile), and (ii) external illumination data; see for instance Léger et al. (2007, http://cdsads.u-strasbg.fr/abs/2007sf2a.conf..592L)

**An important note about the usage of our codes**

We strongly encourage interested scientists to use, and develop further our codes. However, we shall ask them to comply with the following rules:
1. **the first refereed-journal publication of any new user should include both the names of L. Léger and F. Paletou among the co-authors** (with affiliation: *Université Paul Sabatier, Observatoire Midi–Pyrénées, Cnrs, Cnes, IRAP, F-31400 Toulouse*);
2. **the MEDOC service at IAS** *(Orsay, France)* **which distribute these ressources should also be properly aknowledged**;
3. **further publications should be mentioned to** `frederic.paletou@univ-tlse3.fr`

We are grateful to HAO/NCAR (USA), ITA/UiO (Norway) and the IAS (UP-XI, France) who mainly helped fot the development of such an expertise, together with the *priceless* supervision and contribution of L.H. Auer (LANL, USA).

**e-bibliography**

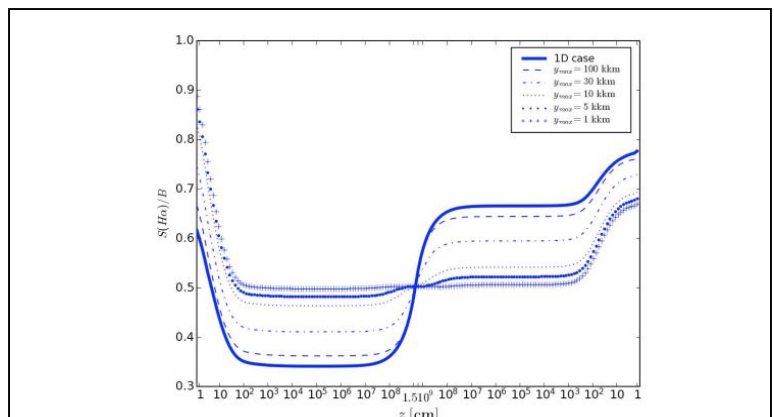

**Fig. 7.** Vertical variations of the H$\alpha$ source function (in units of B) along the symmetry axis of the 2D slab with $z_{max} = 30\,000$ km and different horizontal extensions $y_{max}$ ranging from 1000 km to $\infty$ (1D); the temperature of the atmosphere is 5000 K and the gas pressure $p_g = 1$ dyn cm$^{-2}$. The solid line represents $S(H\alpha)/B$ variations computed in 1D. Note that abscissae give *geometrical* positions computed downwards from the top surface up to mid-slab and then, symmetrically, upward from slab bottom.